# *p*-wave superconductivity in iron-based superconductors

E. F. Talantsev[1,2,*], K. Iida[3,4], T. Ohmura[3], T. Matsumoto[4], W. P. Crump[5,6], N. M. Strickland[5], S. C. Wimbush[5,6] and H. Ikuta[3,4]

[1] M. N. Mikheev Institute of Metal Physics, Ural Branch, Russian Academy of Sciences, 18 S. Kovalevskoy St., Ekaterinburg 620108, Russia

[2] NANOTECH Centre, Ural Federal University, 19 Mira St., Ekaterinburg 620002, Russia

[3] Department of Crystalline Materials Science, Nagoya University, Chikusa-ku, Nagoya 464-8603, Japan

[4] Department of Materials Physics, Nagoya University, Chikusa-ku, Nagoya 464-8603, Japan

[5] Robinson Research Institute, Victoria University of Wellington, 69 Gracefield Road, Lower Hutt 5010, New Zealand

[6] MacDiarmid Institute for Advanced Materials and Nanotechnology, PO Box 33436, Lower Hutt 5046, New Zealand

*Corresponding author: E-mail: evgeny.talantsev@imp.uran.ru

*Abstract*

The possibility of *p*-wave pairing in superconductors has been proposed more than five decades ago, but has not yet been convincingly demonstrated. One difficulty is that some *p*-wave states are thermodynamically indistinguishable from *s*-wave, while others are very similar to *d*-wave states. Here we studied the self-field critical current of NdFeAs(O,F) thin films in order to extract absolute values of the London penetration depth, the superconducting energy gap, and the relative jump in specific heat at the superconducting transition temperature, and find that all the deduced physical parameters strongly indicate that NdFeAs(O,F) is a bulk *p*-wave superconductor. Further investigation revealed that single atomic layer FeSe also shows *p*-wave pairing. Further investigation revealed that single atomic layer FeSe also shows *p*-wave pairing. In an attempt to generalize these findings, we re-examined the whole inventory of superfluid density measurements in iron-based superconductors and show quite generally that single-band weak-coupling *p*-wave superconductivity is exhibited in iron-based superconductors.

The existence of *p*-wave superconductivity was hypothesized more than 50 years ago [1] and the fundamental mechanisms governing *p*-wave superconductivity are well developed in theory. There have however been problems finding a material that convincingly demonstrates *p*-wave superconductivity. The difficulties arise because some *p*-wave states are thermodynamically indistinguishable from *s*-wave states, whilst others would give very similar thermodynamic data to *d*-wave states [2]. Sensitive probes for *p*-wave superconductivity must couple to either the odd parity or the spin part of the pairing. The vast majority of experimental works that have been reported to date concentrate on the latter. In spite of this a material that has bulk *p*-wave pairing remains to be found. $Sr_2RuO_4$ is one of the rare materials in which, for two decades now, *p*-wave superconductivity was thought to exist [3], but recent experiments [4] suggest that it is in all likelihood a *d*-wave superconductor. Thus, there is an on-going experimental search for *p*-wave pairing in new materials [5,6], including induced superconductivity in graphene-based systems [7]. The current status of the search for *p*-wave pairing was recently reviewed in [8,9].

**Description of the problem.** One of the most robust ways of confidently detecting pairing type (i.e., *s*-, *d*-, or *p*-wave) in superconductors is the analysis of the temperature dependence of the superfluid density [10]:

$$\rho_s(T) = \frac{1}{\lambda^2(T)} \quad (1)$$

where $\lambda(T)$ is the London penetration depth. We note that this was the approach used by Hardy *et al.* [11] to demonstrate *d*-wave pairing in high-temperature superconducting cuprates. In Supplementary Fig. 1a we show the normalized superfluid densities, $\rho_s(T/T_c) =$ $((\lambda(0)/\lambda(T/T_c))^2$, for *s*-wave and *d*-wave superconductors and compare them with four possible scenarios of a weak-coupled *p*-wave superconductor in Supplementary Fig. 1b. The analysis of $\rho_s(T)$ for *p*-wave pairing is much more complicated (in comparison with *s*- and *d*-wave) because in this case the gap function is given by [12-14]:

$$\Delta(\hat{\boldsymbol{k}}, T) = \Delta(T) f(\hat{\boldsymbol{k}}, \hat{\boldsymbol{l}}) \qquad (2)$$

where $\Delta$ is the superconducting gap, *k* is the wave vector, and *l* is the gap axis. The electromagnetic response depends on the mutual orientation of the vector potential **A** and the gap axis which for an experiment is just the orientation of the crystallographic axes compared with the direction of the electric current. There are two different *p*-wave pairing states: "axial" where there are two point nodes, and "polar" where there is an equatorial line node. It can be seen from Supplementary Fig. 1 that the only *p*-wave case that is clearly distinguishable from *s*-wave and *d*-wave is polar **A**⊥*l*, which is the only case for which the second derivative of $\rho_s(T/T_c)$ vs. $T/T_c$ has opposite sign to all other scenarios for *s*-, *d*-, and *p*-wave pairing; that is, the temperature dependence of the superfluid density has positive curvature at all temperatures. The shapes of the superfluid densities for *p*-wave polar **A**∥*l* and axial **A**⊥*l* cases are difficult to distinguish from their *s*-wave counterparts, and the *p*-wave axial **A**∥*l* case is also difficult to distinguish from the dirty *d*-wave case.

In spite of these difficulties in the distinguishing of *p*-wave, *s*-wave and *d*-wave cases based on the shape of $\rho_s(T)$, there is still the possibility to make this deduction based on the values of several superconducting parameters deduced from the $\rho_s(T)$ analysis. For instance, Bardeen-Cooper-Schrieffer theory [15] weak-coupling limits for these types of pairing are given in Supplementary Table I [12,13,15-17].

We note that, as mentioned by Gross-Alltag *et al.* [13], only at very particular experimental conditions can the pure polar or pure axial cases of the *p*-wave superconductivity be observed. More likely, as was the case for heavy fermions [12,13], the hybrid cases will be observed in experiments.

In the case of iron-based superconductors (this class of unconventional superconductors includes more than 30 iron based superconductors discovered to date, which have 12 different crystallographic space groups [17,18]), there is an obvious objection to them being *p*-wave superconductors, because Knight shift experiments showed that *p*-wave should be prohibited [16]. We note that consideration of the Knight shift in superconductors started in the early 1960s [19] when it was believed that ferromagnetism is antagonistic to superconductivity. We suggest that a simple extrapolation of theoretical results in regards of the Knight shift obtained for classical BCS superconductors probably is not valid for the newly discovered class of iron-based superconductors.

We stress that there is an exceptional experimental condition under which *p*-wave superconductivity can be uniquely determined from the temperature dependence of the polar $\mathbf{A} \perp \mathbf{l}$ case of $\rho_s(T)$, and thus the lack of experimental studies for confidently detecting *p*-wave pairing is related not just to the fabrication of samples but also choosing an experimental technique for which the polar $\mathbf{A} \perp \mathbf{l}$ orientation can be studied.

If we consider transport current flow in the basal plane *c*-axis oriented *p*-wave superconducting film then this is consistent with the case of polar $\mathbf{A} \perp \mathbf{l}$, which is equatorial line node mode with current flowing in the plane perpendicular to the gap axis. We note that the self-field critical current, $J_c(\mathrm{sf},T)$, in thin superconducting films obeys the relation [20]:

$$J_c(\mathrm{sf},T) = \frac{\phi_0}{4\pi\mu_0} \cdot \frac{ln(\kappa)+0.5}{\lambda^3(T)} = \frac{\phi_0}{4\pi\mu_0} \cdot (ln(\kappa) + 0.5) \cdot \rho_s^{1.5}(T) \qquad (3)$$

where $\phi_0 = 2.067 \times 10^{-15}$ Wb is the magnetic flux quantum, $\mu_0 = 4\pi \times 10^{-7}$ H/m is the magnetic permeability of free space, and $\kappa = \lambda/\xi$ is the Ginzburg-Landau parameter, and thus $J_c(\mathrm{sf},T)$ is proportional to $\rho_s^{1.5}(T)$. In Supplementary Fig. 1c,d we show normalized plots of the temperature dependence of $\rho_s^{1.5}(T/T_c) = ((\lambda(0)/\lambda(T/T_c))^3$ for *s*-, *d*-, and *p*-wave superconductors respectively, where $\lambda(0)$ is the ground-state London penetration depth referring to the value in the limit $T \to 0\ K$.

In this paper, drawing upon previous work [20-22], we studied the self-field critical current density, $J_c(\text{sf},T)$, of NdFeAs(O,F) thin films with the aim of extracting the absolute values of the ground-state London penetration depth, $\lambda(0)$, the ground-state superconducting energy gap, $\Delta(0)$, and the relative jump, $\Delta C/C$, in specific heat at the superconducting transition temperature, $T_c$. Our initial purpose was to make an accurate determination of these superconducting parameters within a multiple *s*-wave gap scenario, due to this being the most widely accepted assumption regarding the superconducting pairing symmetry in iron-based superconductors [17,18].

However, the experimental $J_c(\text{sf},T)$ data as we show below was found to be incompatible with this scenario or even a multi-band *d*-wave scenario. Our analysis revealed that NdFeAs(O,F) is a single-band weak-coupling *p*-wave superconductor with

$$\frac{2\Delta(0)}{k_B T_c} = 5.52 \pm 0.06 \tag{4}$$

where $k_B = 1.381 \times 10^{-23}$ JK$^{-1}$ is the Boltzmann constant. This value is in good agreement with the majority of experimental data on direct measurements of $2\Delta(0)/k_B T_c$ in iron-based superconductors, which is always reported to be in the range from 5 to 6 [16-18].

To further prove our finding and explain why this pairing symmetry was not observed by other techniques, we re-examined available $J_c(\text{sf},T)$ data for thin films of other iron-based superconductors. All *c*-axis oriented thin films for which we re-analyse results herein demonstrate a single band *p*-wave polar $\mathbf{A} \perp \mathbf{l}$ case as our own NdFeAs(O,F) film. These samples are:

1. Single atomic layer FeSe film with $T_c > 100$ K [23];
2. FeSe$_{0.5}$Te$_{0.5}$ thin film with $T_c = 13$ K [24];

3. (Li,Fe)OHFeSe thin film with $T_c$ = 42.2 K [25];
4. Co-doped BaFe$_2$As$_2$ thin film with $T_c$ = 21 K [26];
5. P-doped BaFe$_2$As$_2$ thin film with $T_c$ = 29 K [27];

We also analyse temperature dependent superfluid density, $\rho_s(T)$, for several bulk superconductors:

6. LaFePO single crystal with $T_c$ = 29 K [28];
7. (Li$_{0.84}$Fe$_{0.16}$)OHFe$_{0.98}$Se single crystal with $T_c$ = 42.5 K [29];
8. Rb$_{0.77}$Fe$_{1.61}$Se$_2$ single crystal with $T_c$ = 35.2 K [30];
9. K$_{0.74}$Fe$_{1.66}$Se$_2$ single crystal with $T_c$ = 32.5 K [30];
10. Type-II Weyl semimetal $T_d$-MoTe$_2$ with $T_c$ = 1.48-2.75 K [31].

The latter is not iron-based superconductor, but we show that the formalism of single band *p*-wave superconductivity, we proposed herein, is equally well applied to this superconductor.

We thus found that *p*-wave gap symmetry indeed provides a consistent and reliable description of the whole variety of iron-based superconductors.

**RESULTS**

**The self-field critical current density of thin films**. For a *c*-axis oriented film of an anisotropic superconductor having rectangular cross-section with width 2*a* and thickness 2*b*, the critical current density is given by the following equation [32]:

$$J_c(sf,T) = \frac{\phi_0}{4\pi\mu_0} \cdot \left[ \frac{\ln(\kappa_c)+0.5}{\lambda_{ab}^3(T)} \left( \frac{\lambda_c(T)}{b} \tanh\left(\frac{b}{\lambda_c(T)}\right) \right) + \frac{\ln(\gamma(T)\cdot\kappa_c)+0.5}{\sqrt{\gamma(T)}\cdot\lambda_{ab}^3(T)} \left( \frac{\lambda_{ab}(T)}{a} \tanh\left(\frac{a}{\lambda_{ab}(T)}\right) \right) \right] \quad (5)$$

where $\lambda_{ab}(T)$ and $\lambda_c(T)$ are the in-plane and out-of-plane London penetration depths respectively, $\kappa_c = \lambda_{ab}(T)/\xi_{ab}(T)$ and the electron mass anisotropy $\gamma(T) = \lambda_c(T)/\lambda_{ab}(T)$. For isotropic superconductors $\gamma(T) \equiv 1$ and isotropic Ginzburg-Landau parameter, $\kappa = \lambda(T)/\xi(T)$, replaces $\kappa_c$ in Eq. 5.

Although it is well established [33] that $\gamma(T)$ in iron-based superconductors is temperature dependent, in the case of thin films ($b < \lambda_{ab}(0) \ll a$), Eq. 5 reduces to:

$$J_c(sf,T) = \frac{\phi_0}{4\pi\mu_0} \cdot \left[\frac{ln(\kappa_c)+0.5}{\lambda_{ab}^3(T)} + \frac{ln(\kappa_c)+ln(\gamma(T))+0.5}{\sqrt{\gamma(T)}\cdot\lambda_{ab}^2(T)\cdot a}\right] \cong \frac{\phi_0}{4\pi\mu_0} \cdot \left[\frac{ln(\kappa_c)+0.5}{\lambda_{ab}^3(T)}\right] \quad (6)$$

which is independent of $\gamma(T)$.

Based on this, in our analysis we use the ground state electron mass anisotropy $\gamma(0) = \lambda_c(0)/\lambda_{ab}(0)$ which was taken from independent experimental reports and the basic equation for analysis of $J_c(sf,T)$ was the following:

$$J_c(sf,T) = \frac{\phi_0}{4\pi\mu_0} \cdot \left[\frac{ln(\kappa_c)+0.5}{\lambda_{ab}^3(T)}\left(\frac{\lambda_c(T)}{b}\tanh\left(\frac{b}{\lambda_c(T)}\right)\right) + \frac{ln(\gamma(0)\cdot\kappa_c)+0.5}{\sqrt{\gamma(0)}\cdot\lambda_{ab}^3(T)}\left(\frac{\lambda_{ab}(T)}{a}\tanh\left(\frac{a}{\lambda_{ab}(T)}\right)\right)\right]. \quad (7)$$

***NdFeAs(O,F) thin films.*** We have prepared thin films with two thicknesses of $2b$ = 30 and 90 nm. The Ginzburg-Landau parameter $\kappa_c$ = 90 for NdFeAs(OF) [17,18,34,35] and its electron mass anisotropy $\gamma$ = 5 [36]. Processed experimental $J_c(sf,T)$ data for a NdFeAs(OF) thin film (bridge width $2a$ = 9 μm, film thickness $2b$ = 90 nm) is shown in Fig. 1(a) together with the absolute values of $\lambda(T)$ calculated by numerical solution of Eq. 7.

In Fig. 1(a) we also show the value of the ground-state London penetration depth $\lambda(0)$ = 195 nm measured by μSR for NdFeAsO$_{0.85}$ as reported by Khasanov *et al.* [37]. In Fig. 1(b) we have undertaken a manual scaling of $\rho_s^{1.5}(T)$ to the experimental $J_c(sf,T)$ data for weak coupled *s*-wave, *d*-wave, *p*-wave axial **A**||***l***, and *p*-wave polar **A**⊥***l*** cases. It can be seen that only the latter provides a reasonable fit.

To deduce the fundamental superconducting parameters of the NdFeAs(O,F) thin film from the $J_c(sf,T)$ data we employ the general approach of BCS theory [15], in which the thermodynamic properties of a superconductor are derived from the superconducting energy gap, $\Delta(T)$. We used the temperature-dependent superconducting gap $\Delta(T)$ equation for the *p*- wave polar **A**⊥*l* case given by Gross-Alltag *et al.* [12,13] (which allows for variation in the coupling strength):

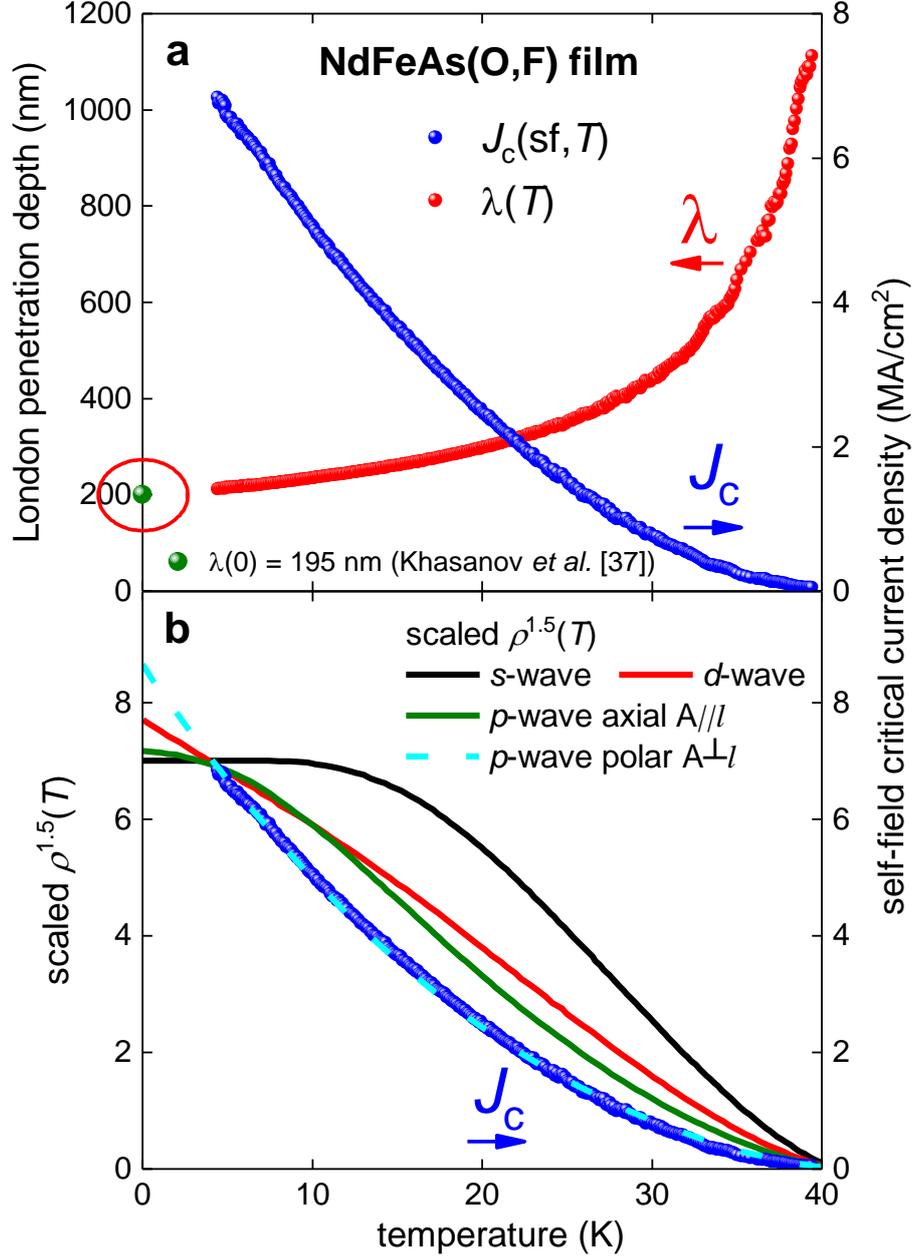

**Figure 1.** (a) Experimental $J_c(\text{sf},T)$ and $\lambda(T)$ calculated from Eq. 7 for a NdFeAs(OF) thin film; $\kappa_c = 90$ was used [17,18,34,35]. Green data point indicates $\lambda(0) = 195$ nm deduced by Khasanov *et al.* [37] for NdFeAsO$_{0.85}$. (b) Scaling of $\rho_s^{1.5}(T)$ for *s*-, *d*- and *p*-wave pairing to the experimental $J_c(\text{sf},T)$ data.

$$\Delta(T) = \Delta(0) \tanh\left( \frac{\pi k_B T_c}{\Delta(0)} \sqrt{\eta \left(\frac{\Delta C}{C}\right)\left(\frac{T_c}{T} - 1\right)} \right) \quad (8)$$

with

$$\eta = \frac{2}{3}\frac{1}{\int_0^1 f^2(x)dx} \tag{9}$$

where, for polar $p$-wave:

$$f(x) = x \tag{10}$$

and for axial $p$-wave:

$$f(x) = \sqrt{1-x^2}. \tag{11}$$

The equation for $\lambda(T)$ was also given by Gross-Alltag *et al.* [12,13]:

$$\lambda(T) = \frac{\lambda(0)}{\sqrt{1-\frac{3}{4k_B T}\int_0^1 \frac{1-x^2}{2}\left[\int_0^\infty \frac{d\varepsilon}{\cosh^2\left(\frac{\sqrt{\varepsilon^2+\Delta^2(T)f^2(x)}}{2k_B T_C}\right)}\right]dx}} \tag{12}$$

By substituting Eqs. 8-12 in Eq. 7 for thick samples, or by using Eqs. 3,6 for thin samples for which the film thickness, $2b < \lambda(0)$, one can fit the experimental $J_c(\text{sf},T)$ data to the model and deduce $\lambda(0)$, $\Delta(0)$, $\Delta C/C$ and $T_c$ as free-fitting parameters. To help experimentalists use our model to infer $\lambda(0)$, $\Delta(0)$, $\Delta C/C$ and $T_c$ parameters from measured $J_c(\text{sf},T)$ data (which is not a trivial mathematical task), we have made our fitting code available online [38].

The result of the fit is shown in Fig. 2 and the parameters derived from the fit are found to be in good agreement with weak-coupling values predicted by BCS theory given by Gross-Alltag *et al.* [12,13]. For instance, the deduced $\Delta C/C = 0.80 \pm 0.01$ and $2\Delta(0)/k_B T_c = 5.52 \pm 0.06$ compare well with the predicted BCS weak-coupling values for polar orientation of 0.792 and 4.924, respectively (Supplementary Table I). In Fig. 1 we also show the value of the ground-state London penetration depth $\lambda(0) = 195$ nm measured by μSR for NdFeAsO$_{0.85}$ as reported by Khasanov *et al.* [37].

A similar BCS ratio of $2\Delta(0)/k_B T_c = 5.0\text{-}5.7$ was found in the related compound Sm$_{1-x}$Th$_x$OFeAs reported by Kuzmicheva *et al.* [39]. The weak-coupling scenario was also experimentally found in the related compound LaFeAsO$_{0.9}$F$_{0.1}$ [40].

The deduced ground-state London penetration depth λ(0) = 198.2 ± 0.1 nm is also in very good agreement with independent measurements showing λ(0) = 195-200 nm [37]. These results strongly support the conclusion that NdFeAs(O,F) is a *p*-wave superconductor.

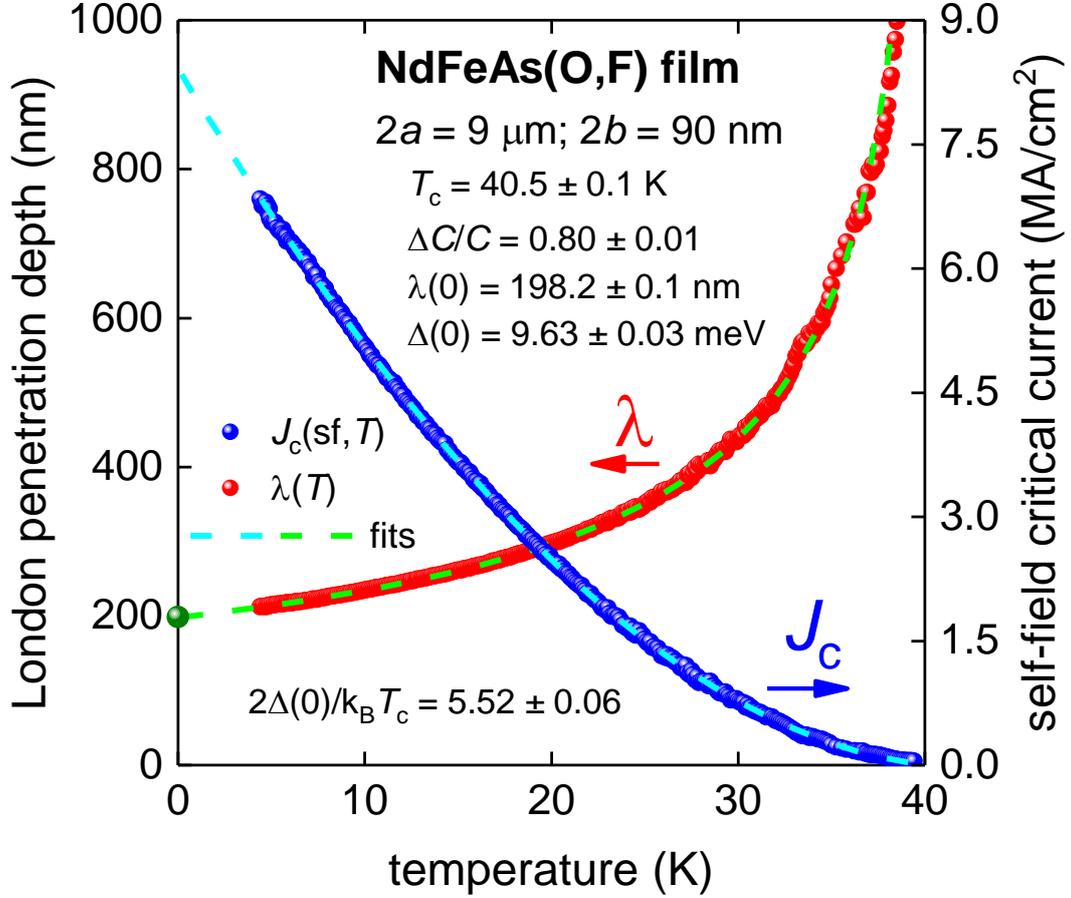

**Figure 2.** BCS fits to the experimental $J_c$(sf,$T$) data and $\lambda(T)$ calculated from Eq. 7 for a NdFeAs(O,F) thin film assuming a *p*-wave polar **A**⊥*l* model (Eqs. 7-12) and $\kappa_c$ = 90 [17,18,34,35]. Green data point indicates $\lambda(0)$ = 195 nm deduced by Khasanov *et al.* [37] for NdFeAsO$_{0.85}$. Derived parameters are: $T_c$ = 40.5 ± 0.5 K, $\Delta(0)$ = 9.63 ± 0.03 meV, $\Delta C/C$ = 0.80 ± 0.01, $\lambda(0)$ = 198.2 ± 0.1 nm, $2\Delta(0)/k_B T_c$ = 5.52 ± 0.06. Fit quality is $R$ = 0.99995.

*FeSe single atomic layer film.* To support our finding that some iron-based superconductors have *p*-wave pairing symmetry we performed a search for experimental $J_c$(sf,$T$) datasets for these materials. In Figure 3 we show $J_c$(sf,$T$) and fit to Eqs. 7 for the milestone report about FeSe single atomic layer sample with record transition temperature, $T_c \gtrsim$ 109 K, reported by Ge *et al.* [23].

To make this fit we made the assumption that the in-plane Ginzburg-Landau parameter $\kappa_c$ = 72 does not change from its bulk [41] and other single atomic layer film [42,43] values. The deduced $\lambda_{ab}(0)$ = 167 nm is in good agreement with this assumption, taking into account that $\xi_{ab}(0)$ = 2.4 nm [43]. The fit to the *p*-wave model (Eqs. 7-12) revealed that $2\Delta(0)/k_BT_c$ = 4.9 ± 0.6 which is equal to the *p*-wave weak-coupling limit (Supplementary Table I) and more data are required to deduce $\Delta C/C$ with greater accuracy.

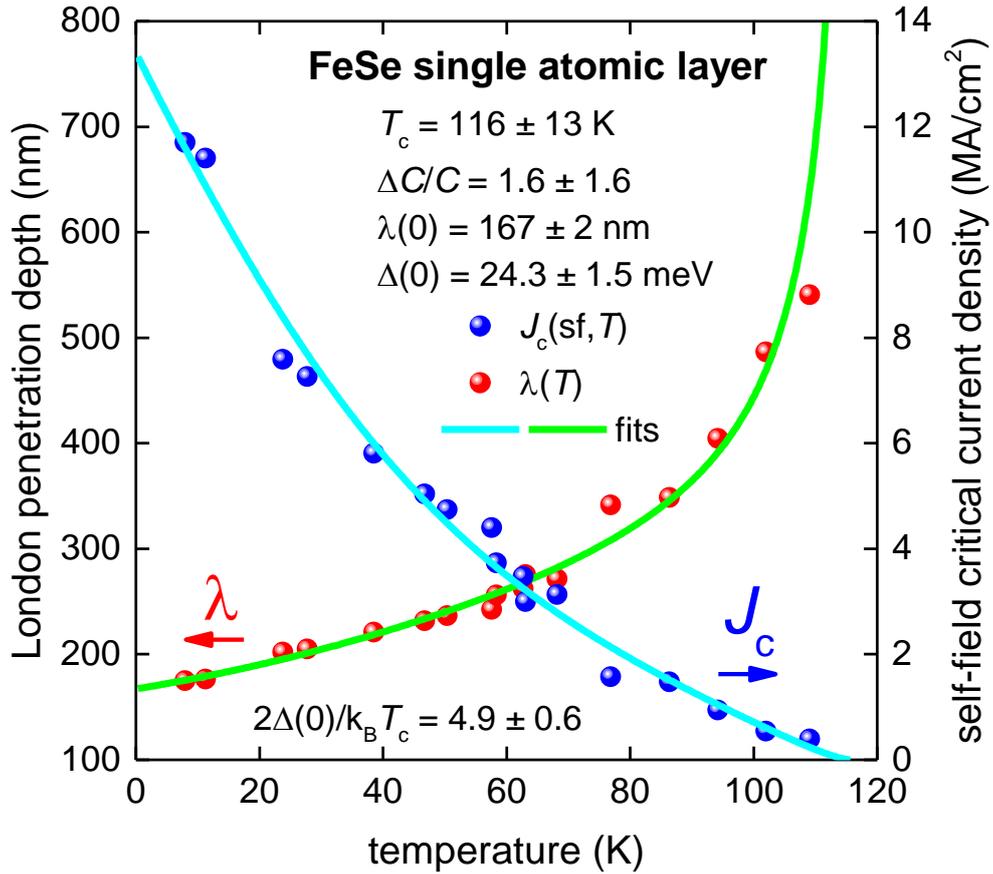

**Figure 3.** BCS fits to the experimental $J_c$(sf,$T$) data [23] and $\lambda(T)$ calculated from Eq. 7 for a single atomic layer FeSe film assuming a *p*-wave polar $\mathbf{A}\perp\mathbf{l}$ model, and $\kappa_c$ = 72. Derived parameters are: $T_c$ = 116 ± 13 K, $\Delta(0)$ = 24.3 ± 1.5 meV, $\Delta C/C$ = 1.6 ± 1.6, $\lambda(0)$ = 167 ± 2 nm, $2\Delta(0)/k_BT_c$ = 4.9 ± 0.6. Fit quality is $R$ = 0.8564.

***FeSe$_{0.5}$Te$_{0.5}$ thin film.*** The next example found in the literature is an FeSe$_{0.5}$Te$_{0.5}$ thin film (2*a* = 800 nm, 2*b* = 100 nm) where the raw $J_c$(sf,$T$) data from Nappi *et al.* [24] is shown in Fig. 4.

To make a fit of $J_c(sf,T)$ using Eqs. 7-12, we used a Ginzburg-Landau parameter $\kappa_c = 180$ [33,44] and electron mass anisotropy $\gamma = 2.5$ [45,46].

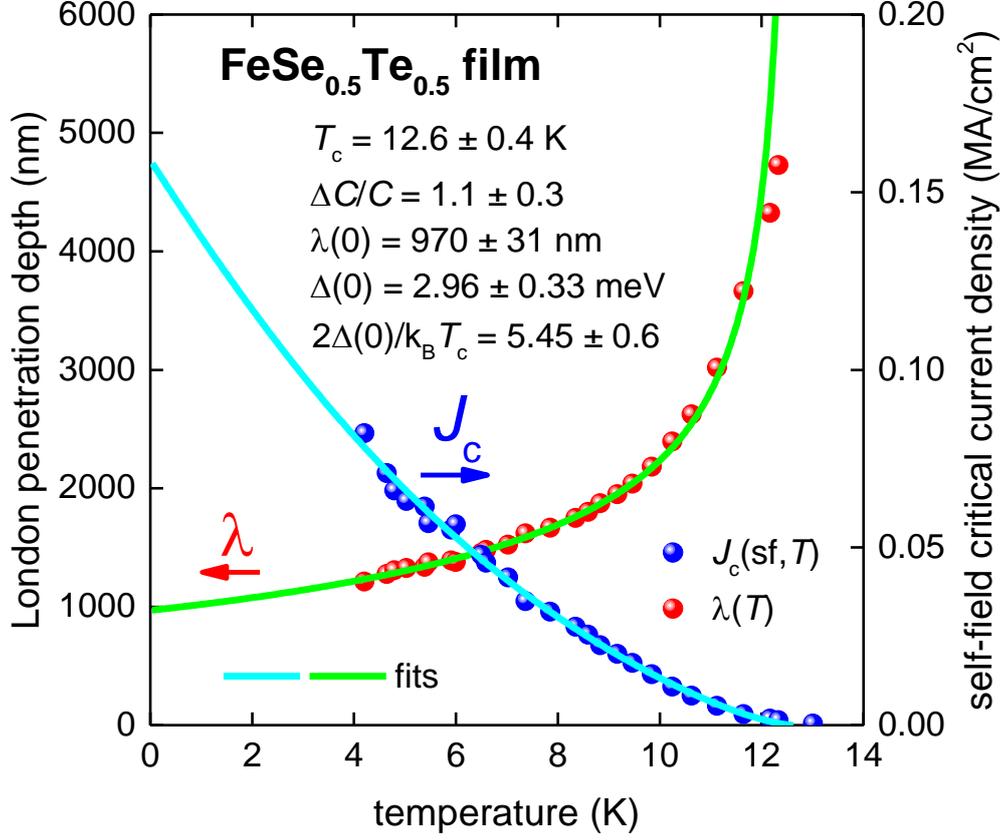

**Figure 4.** BCS fits to the experimental $J_c(sf,T)$ data [24] and $\lambda(T)$ calculated from Eq. 7 for an FeSe$_{0.5}$Te$_{0.5}$ thin film ($2a = 800$ nm, $2b = 100$ nm) assuming a $p$-wave polar $\mathbf{A}\perp l$ model, $\kappa_c = 180$ and $\gamma = 2.5$. Derived parameters are: $T_c = 12.6 \pm 0.4$ K, $\Delta(0) = 2.96 \pm 0.33$ meV, $\Delta C/C = 1.1 \pm 0.3$, $\lambda(0) = 970 \pm 31$ nm, $2\Delta(0)/k_B T_c = 5.45 \pm 0.6$. Fit quality is $R = 0.9711$.

As can be seen, the fit matches excellently with the weak-coupling polar $\mathbf{A}\perp l$ $p$-wave case. We note that the derived $\lambda(0) = 970 \pm 31$ nm is larger than the value reported by Bendele *et al.*, $\lambda(0) = 492$ nm [45]. We expect that this difference is related to some information mentioned by Nappi *et al.* [24], that during the preparation of the transport current bridge, the transition temperature of the film was reduced. We hypothesize that there was some minor damage caused to the current bridge edges. Based on this, the dissipation-free transport current is flowing along a narrower bridge, and thus the actual $J_c(sf,T)$ will be higher than that calculated based on the nominal sample width $2a$. Lower temperature data would of course be desirable to support our case for a $p$-wave scenario more strongly.

*(Li,Fe)OHFeSe thin film*. The next thin film presented here is (Li,Fe)OHFeSe ($2a = 50$ μm, $2b = 20$ nm) where the raw $J_c$(sf,$T$) data reported by Huang *et al.* [25] is shown in Fig. 5. For a fit of $J_c$(sf,$T$) using Eqs. 7-12, we take into account measurements of the in-plane coherence length $\xi_{ab}(0) = 2.0$ nm [47,48] and $\lambda_{ab}(0) = 280$ nm [29], which give the Ginzburg-Landau parameter as $\kappa_c = 140$. The electron mass anisotropy for this compound is $\gamma = 10$ [29].

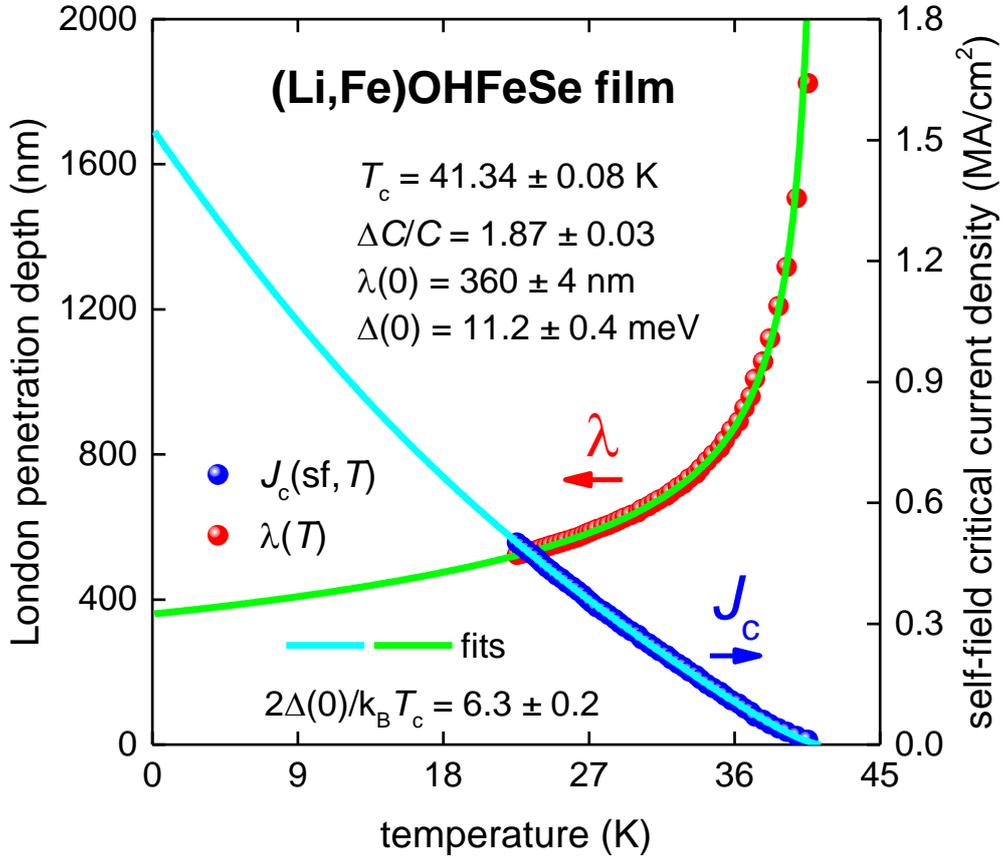

**Figure 5.** BCS fits to the experimental $J_c$(sf,$T$) data [25] and $\lambda(T)$ calculated from Eqs. 7-12 for a (Li,Fe)OHFeSe thin film ($2a = 50$ μm, $2b = 20$ nm) assuming a $p$-wave polar $\mathbf{A} \perp \mathbf{l}$ model, $\kappa_c = 140$ and $\gamma = 10$. Derived parameters are: $T_c = 41.34 \pm 0.08$ K, $\Delta(0) = 11.2 \pm 0.4$ meV, $\Delta C/C = 1.87 \pm 0.03$, $\lambda(0) = 360 \pm 4$ nm, $2\Delta(0)/k_B T_c = 6.3 \pm 0.2$. Fit quality is $R = 0.9997$.

Despite the lack of low-temperature data points, the deduced $\lambda(0) = 360 \pm 4$ nm is in reasonable agreement with the value $\lambda(0) = 280$ nm measured in a $(Li_{0.84}Fe_{0.16})OHFe_{0.98}Se$ single crystal by μSR experiments [29].

The deduced ratio of $2\Delta(0)/k_BT_c = 6.3 \pm 0.2$ along with $\Delta C/C = 1.87 \pm 0.03$ together show that (Li,Fe)OHFeSe is likely a moderately strongly coupled *p*-wave superconductor. Analysis of the superfluid density measured by μSR on bulk $(Li_{0.84}Fe_{0.16})OHFe_{0.98}Se$ single crystals also reveals similar values of $2\Delta(0)/k_BT_c$ and $\Delta C/C$ as those derived using $J_c(sf,T)$ data. This analysis is presented in the Supplementary Information Section S1.

***Co-doped BaFe$_2$As$_2$ thin film.*** Now we consider the most studied but perhaps least understood and most puzzling iron-based superconductor, $BaFe_2As_2$. This compound can be made to superconductor by substituting on different atomic sites. One of the most representative examples of the self-field critical current density in Co-doped $BaFe_2As_2$ was reported by Tarantini *et al.* [26]. Raw $J_c(sf,T)$ data [26] for the sample with $2a = 40$ μm, $2b = 350$ nm is shown in Figs. 6 and 7. To fit the $J_c(sf,T)$ dataset to Eqs. 7-12, we take into account the Ginzburg-Landau parameter as $\kappa_c = 66$ [49], and the electron mass anisotropy for this compound as $\gamma = 1.5$ [50].

There is a widely accepted view that this compound is a two-band *s*-wave superconductor [17,18]. In the case of a two-band superconductor that has completely decoupled bands, $J_c(sf,T)$ can be written in the form [21,22]:

$$J_c(sf,T)_{\text{total}} = J_c(sf,T)_{\text{band1}} + J_c(sf,T)_{\text{band2}} \tag{12}$$

where $J_c(sf,T)$ for each band is as described by Eq. 3 with separate $\lambda(0)$, $\Delta(0)$, $\Delta C/C$ and $T_c$ values and all eight parameters may be used as free-fitting parameters. The raw $J_c(sf,T)$ dataset measured by Tarantini *et al.* [26] was sufficiently rich that we were able to fit using all eight parameters. For *s*-wave superconductors the gap equation, $\Delta(T)$, is given by Eq. 8 with $\eta = 2/3$, and $\lambda(T)$ is given by [12,13]:

$$\lambda(T) = \frac{\lambda(0)}{\sqrt{1 - \frac{1}{2 \cdot k_B \cdot T} \cdot \int_0^\infty \cosh^{-2}\left(\frac{\sqrt{\varepsilon^2 + \Delta^2(T)}}{2 \cdot k_B \cdot T}\right) d\varepsilon}} \tag{13}$$

More details and examples of application of this *s*-wave weakly-coupled bands model can be found elsewhere [21,22].

The fit to this model is shown in Fig. 6. The fit quality is very high, $R = 0.9993$, and the deduced parameters for both bands agree well with other reports. The downside of this fit, as well as all previously applied two-band *s*-wave models, is that the deduced parameters are at times lower than the BCS weak-coupling limits.

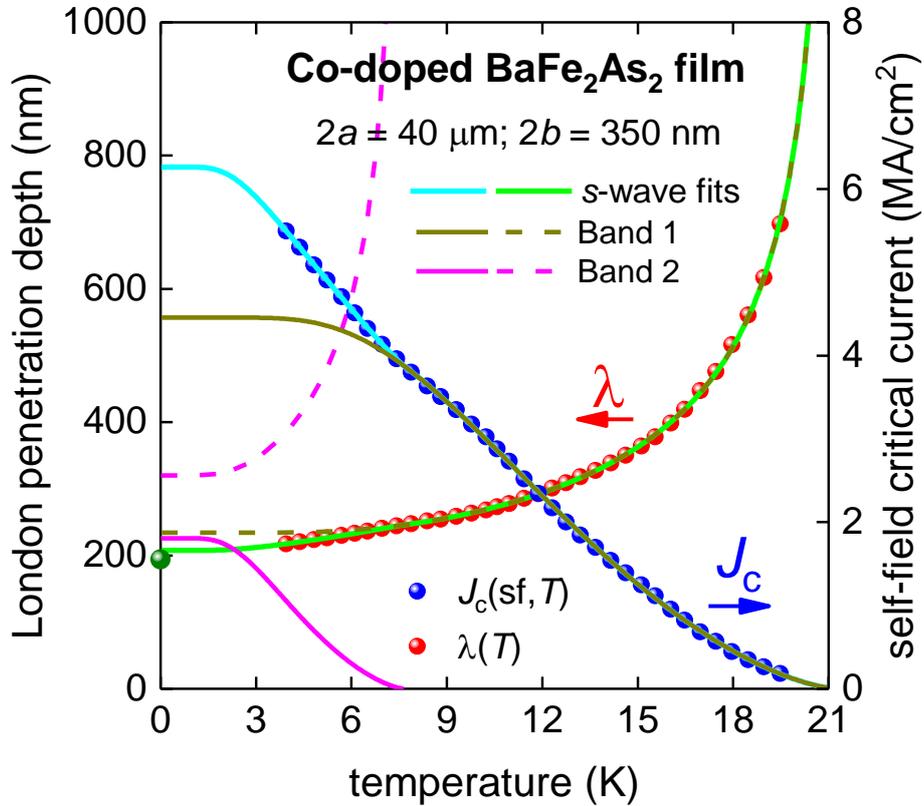

**Figure 6.** BCS fits to the experimental $J_c(\text{sf},T)$ data [26] and $\lambda(T)$ calculated from Eq. 7 for a Co-doped BaFe$_2$As$_2$ thin film ($2a = 40$ μm, $2b = 350$ nm) assuming a two-band *s*-wave model, $\kappa_c = 66$ and $\gamma = 1.5$. Derived parameters are: $T_{c1} = 21.24 \pm 0.16$ K, $\Delta_1(0) = 2.74 \pm 0.05$ meV, $\Delta C_1/C_1 = 0.93 \pm 0.05$, $\lambda_1(0) = 234.8 \pm 0.8$ nm, $2\Delta_1(0)/k_B T_{c1} = 2.99 \pm 0.05$, $T_{c2} = 7.6 \pm 0.2$ K, $\Delta_2(0) = 0.92 \pm 0.18$ meV, $\Delta C_2/C_2 = 1.0 \pm 0.2$, $\lambda_2(0) = 318 \pm 23$ nm, $2\Delta_2(0)/k_B T_{c2} = 2.8 \pm 0.5$. Fit quality is $R = 0.9993$. Green ball is $\lambda(0) = 190$ nm for Co-doped BaFe$_2$As$_2$ [49].

For instance, $\frac{2\Delta(0)}{k_B T_c} < 3$ for both bands as compared with the BCS weak-coupling limit of $\frac{2\Delta(0)}{k_B T_c} = 3.53$, and $\left.\frac{\Delta C}{C}\right|_{T \sim T_c} \lesssim 1$ for both bands as compared with the BCS weak-coupling limit of $\left.\frac{\Delta C}{C}\right|_{T \sim T_c} = 1.43$.

The fit of the same $J_c(\text{sf},T)$ dataset to a single-band polar $\mathbf{A} \perp \mathbf{l}$ $p$-wave model is presented in Fig. 7, where the deduced $\lambda(0) = 198.0 \pm 0.2$ nm is in a good agreement with the reported $\lambda(0) = 190$ nm for cobalt-doped Ba-122 compounds [49]. The other deduced parameters show that this compound has moderately strong coupling.

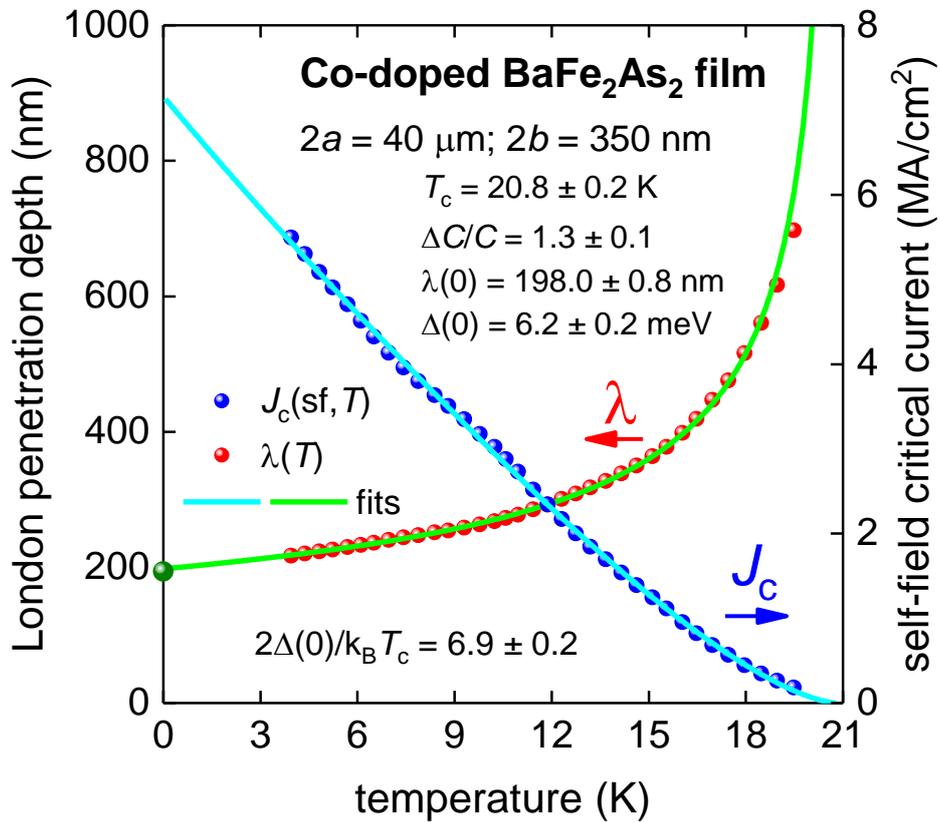

**Figure 7.** BCS fits to the experimental $J_c(\text{sf},T)$ data [26] and $\lambda(T)$ calculated from Eqs. 7-12 for a Co-doped BaFe$_2$As$_2$ thin film ($2a = 40$ μm, $2b = 350$ nm) assuming a $p$-wave polar $\mathbf{A} \perp \mathbf{l}$ model. Derived parameters are: $T_c = 20.8 \pm 0.2$ K, $\Delta(0) = 6.2 \pm 0.2$ meV, $\Delta C/C = 1.3 \pm 0.1$, $\lambda(0) = 198.0 \pm 0.8$ nm, $2\Delta(0)/k_B T_c = 6.9 \pm 0.2$. Fit quality is $R = 0.9979$. Green ball is $\lambda(0) = 190$ nm for Co-doped BaFe$_2$As$_2$ [49].

The significant advantage of this approach is that the fit has only four free-fitting parameters compared with eight for the two-band $s$-wave model. The additional four parameters for the two-band *s*-wave model give a remarkably insignificant improvement in the fit quality ($R = 0.9993$ compared to $R = 0.9979$), while dramatically increasing the mutual interdependency of the fit parameters. Similar arguments apply to the more exotic order parameter symmetries proposed, such as three-band models or *s+is* chiral symmetry models [16-18].

We consider that a good reason must be presented for requiring a more complex model than is needed to adequately explain the experimental data [51,52].

Also, it should be stressed that an unavoidable weakness of all multi-band models, ignoring the overwhelmingly large number of free-fitting parameters within these models, is that at least for one band (or, for two bands in the three-bands models) the ratio of the superconducting energy gap to the transition temperature is several times lower than the lowest value allowed within the most established theory of superconductivity, which is BCS [15]:

$$\frac{2\Delta(0)}{k_B T_c} \ll 3.53 \qquad (14)$$

Thus, in this paper, we present a model which is:

1. framed within the standard BCS single band theory.
2. provides superconducting parameters within weak-coupling BCS limits.

This means that our model is based on a minimal set of physical assumptions and provides values for several structurally different superconductors within the simplest weak-coupling BCS limits. In the next Section we make direct demonstration how experimental data can be processed within $s_\pm$ and *p*-wave models, for which we chose bulk $T_d$-MoTe$_2$ superconductor for which Guguchia *et al.* [31] performed $\rho_s(T)$ data fits to several conventional models, including *s*-wave, *d*-wave, and $s_\pm$ and $s_{++}$ models. And thus, this makes possible to compare our approach with ones proposed previously.

***Bulk sample of Type-II Weyl semimetal Td-MoTe₂***. Guguchia *et al.* [31] reported the temperature dependent $\rho_s(T)$ subjected to high pressure and performed data analysis for single band *s*-wave, *d*-wave, and $s_{\pm}$-wave models in their Fig. 4. In our Fig. 8 we show raw $\rho_s(T)$ data fitted to polar $\mathbf{A}\|\mathbf{l}$ model. It can be seen that fits have very high quality.

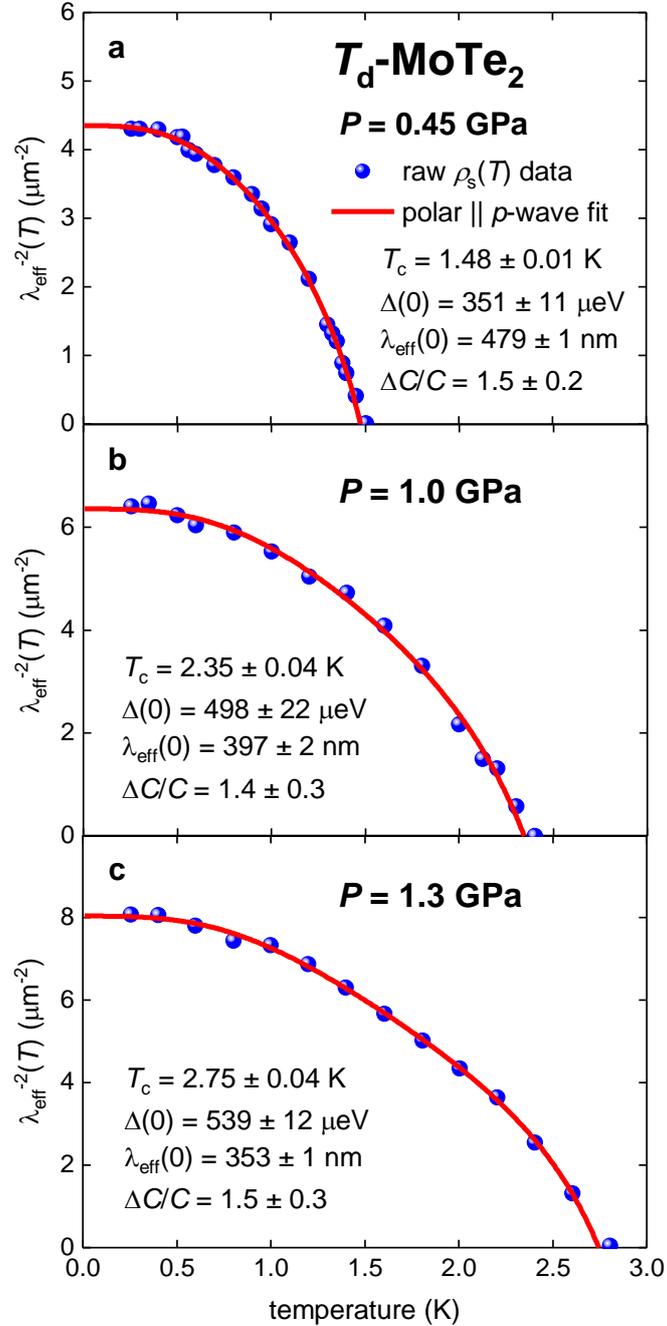

**Figure 8.** BCS fits of $\rho_s(T)$ data for bulk type-II Weyl semimetal $T_d$-MoTe₂ [31] measured at applied field of $B = 20$ mT to single-band *p*-wave polar $\mathbf{A}\|\mathbf{l}$ model (Eqs. 8-12). (a) fit quality is $R = 0.9898$; (b) $R = 0.9650$; (b) $R = 0.9873$.

In Fig. 9(a) we show evolution of $\frac{2\Delta_i(0)}{k_B T_c}$ ratios vs applied pressure deduced by Guguchia *et al.* [31] within $s_\pm$-wave model. It can be seen that for Band 1 deduced ratio is in 2-3.5 times lower than the lowest value of 3.53 allowed by BCS theory for *s*-wave pairing symmetry.

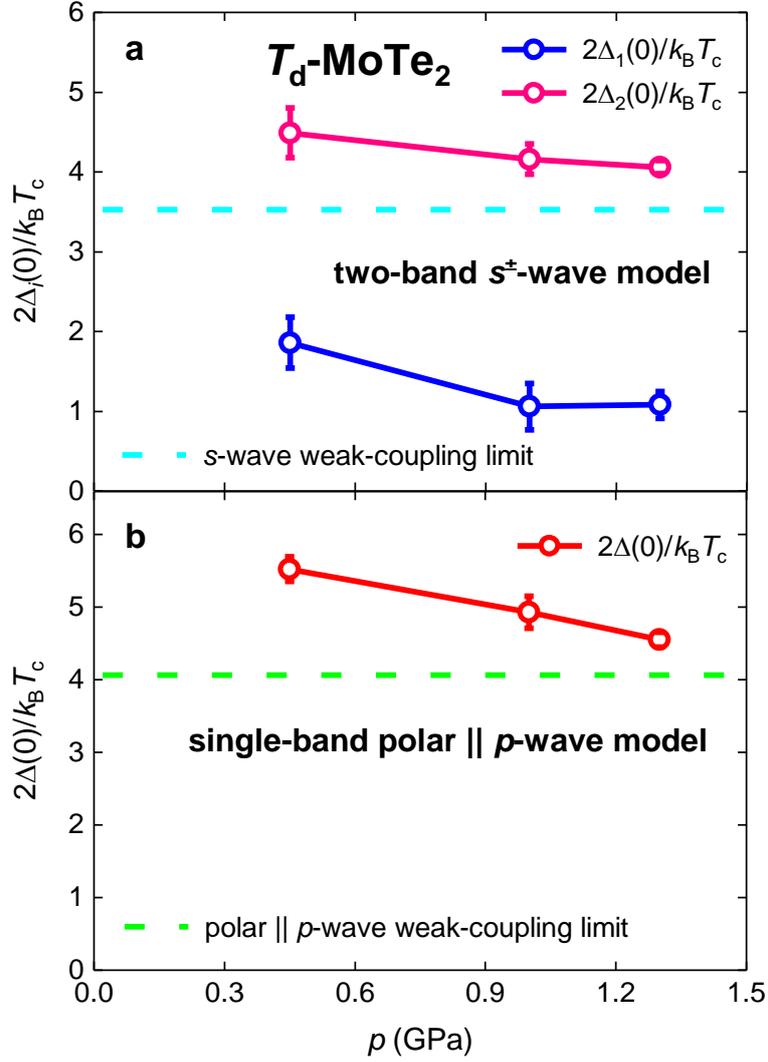

**Figure 9.** Deduced $\frac{2\Delta(0)}{k_B T_c}$ values for bulk type-II Weyl semimetal $T_d$-MoTe$_2$ measured at applied field of $B = 20$ mT [31]. (a) two-band $s_\pm$-wave model. (b) single-band *p*-wave polar **A**||*l* model.

In Fig. 9(b) we show $\frac{2\Delta(0)}{k_B T_c}$ evolution vs applied pressure for the value deduced by applying single-band *p*-wave polar **A**||*l* model. Deduced ratios for this model demonstrate lower uncertainties in comparison with $s_\pm$ model, and also ones show that $T_d$-MoTe$_2$ is moderately strong-coupling superconductor, for which coupling strength is linearly reducing towards weak-coupling limit of 4.06 (for this symmetry) while pressure is increased. This behavior is expected for this quasi-2D material as a direct consequence of the increase in interlayer coupling for 2D-nanosheets while applied pressure is increasing.

This example demonstrates that *p*-wave pairing symmetry perhaps is common feature for many unconventional superconductors.

***P-doped BaFe$_2$As$_2$ thin film.*** Kurth *et al.* [27] reported the self-field critical current density for isovalently P-doped BaFe$_2$As$_2$ (Ba-122) single crystalline thin films deposited on MgO (001) substrates by molecular beam epitaxy. The film dimensions were $2a = 40$ μm, and $2b = 107$ nm. In Fig. 10 we show a fit of $J_c(sf,T)$ to Eqs. 7-12 using $\gamma = 2.6$ [53] and $\kappa = 93$ [49].

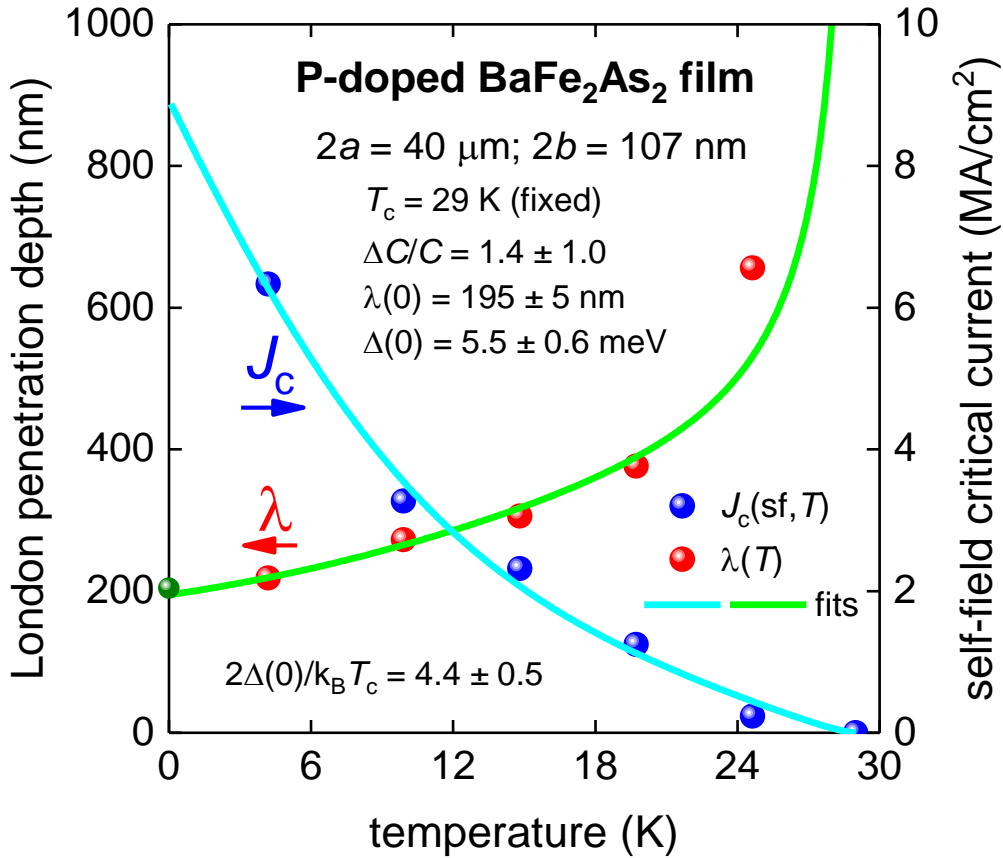

**Figure 10.** BCS fits to the experimental $J_c(sf,T)$ data [27] and $\lambda(T)$ calculated from Eqs. 7-12 for a P-doped BaFe$_2$As$_2$ thin film ($2a = 40$ μm, $2b = 107$ nm) assuming a *p*-wave polar **A**⊥*l* model, $\kappa = 93$ and $\gamma = 2.6$. $T_c$ was fixed at 29 K. Derived parameters are: $\Delta(0) = 5.5 \pm 0.6$ meV, $\Delta C/C = 1.4 \pm 1.0$, $\lambda(0) = 195 \pm 5$ nm, $2\Delta(0)/k_BT_c = 4.4 \pm 0.5$. Fit quality is $R = 0.486$. Green ball is $\lambda(0) = 200$ nm for P-doped BaFe$_2$As$_2$ [49].

Due to the experimental $J_c(sf,T)$ dataset being limited to five data points, we fixed the transition temperature for the fit to $T_c$ = 29 K. The deduced value of $\lambda(0) = 195 \pm 5$ nm is in excellent agreement with the reported $\lambda(0) = 200$ nm for phosphorus-doped Ba-122 compounds [49]. A richer experimental $J_c(sf,T)$ dataset would be beneficial for more accurate determination of the other superconducting parameters.

*LaFePO bulk single crystal: polar* $\mathbf{A}\perp l$. This was observed by a high resolution susceptometer based on a self-resonant tunnel diode circuit by Fletcher *et al.* [28]. In Fig. 11 we show the raw data for their LaFePO Sample #1 with a fit to a *p*-wave $\rho_s(T)$ polar $\mathbf{A}\perp l$ model. We fixed the $T_c$ to the experimental value of 5.45 K. The fit is excellent across a very wide temperature range. All the deduced values are in excellent agreement with the weak-coupling limits of the *p*-wave polar $\mathbf{A}\perp l$ case.

The temperature dependent superfluid density, $\rho_s(T)$, in iron-based superconducting crystals has also been measured directly using muon-spin rotation (µSR) spectroscopy. For most iron-based superconductors reported in the literature, there is again the consistent observation that *p*-wave pairing symmetry exists in these materials. This analysis is given in the Supplementary Information.

As we show in this paper, experimental data for many iron-based superconductors clearly shows that *p*-wave superconductivity is surprisingly often observed in these materials.

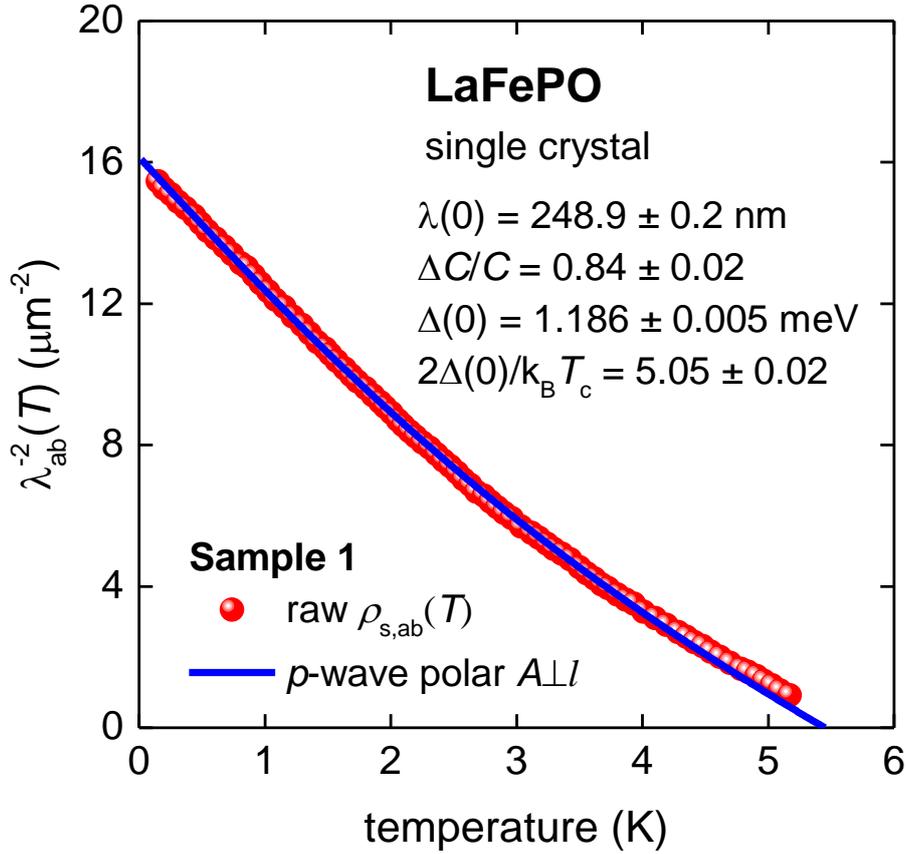

**Figure 11.** BCS fits to the experimental $\rho_{s,ab}(T)$ data [28] for a LaFePO sample assuming a $p$-wave polar $\mathbf{A}\perp l$ case. Derived parameters are: $T_c$ = 5.45 K (fixed), $\Delta(0)$ = 1.186 ± 0.005 meV, $\Delta C/C$ = 0.84 ± 0.02, $\lambda(0)$ = 248.9 ± 0.2 nm, $2\Delta(0)/k_B T_c$ = 5.05 ± 0.02. Fit quality is $R$ = 0.9996.

## SUMMARY

Analysis of self-field critical current data and superfluid density data obtained on a wide variety of iron-based superconductors using $p$-wave models find superconducting parameters (ground-state penetration depth, superconducting gap polar $\mathbf{A}\perp l$ magnitude, and specific heat jump at the transition temperature) that are more consistent under a $p$-wave model compared with the generally-accepted $s$-wave model. Also, observation of the polar $\mathbf{A}\perp l$ model (where the shape is completely different to both $s$- and $d$-wave models) in both the self-field critical current data and superfluid density data strongly indicates the existence of $p$-wave pairing in these iron-based superconductors.

**METHODS**

Superconducting NdFeAs(O,F) thin films were prepared by molecular beam epitaxy. First, the parent compound NdFeAsO was grown on MgO(001) at 800°C, followed by the deposition of a NdOF over-layer at the same temperature, from which fluorine diffused into the NdFeAsO layer [47,48]. Reflection high-energy electron diffraction confirmed the epitaxial growth of NdFeAsO as well as NdOF with smooth surfaces. Since NdOF is an isulator, the NdOF cap layer was removed by ion-beam etching for transport measurements. The NdFeAs(O,F) film was photolithographically patterned and ion-beam etched to fabricate bridges of 9 μm and 20 μm width and 1 and 2 mm in length.

For measurements of $J_c$(sf,$T$), a new system was built based on the Quantum Design Physical Property Measurement System. The new system adopts several parts of the system described elsewhere [56]. A detailed account of the design and operational performance of the new system is given in Ref. 57. The system is capable of supplying transport currents up to 30 A while maintaining a sample temperature of $T = 2.0 \pm 0.1$ K, and currents up to 200 A at higher sample temperatures.

**Acknowledgements**

The authors thank Prof. Jeffery L. Tallon (Victoria University of Wellington, New Zealand) and Prof. Christian Bernhard (University of Fribourg, Switzerland) for helpful discussions, and also for reading and commenting on the manuscript.

EFT is grateful for financial support provided by the state assignment of Minobrnauki of Russia (theme "Pressure" No. AAAA-A18-118020190104-3) and by Act 211 of the Government of the Russian Federation, contract No. 02.A03.21.0006. KI and HI acknowledge support by the Japan Society for the Promotion of Science (JSPS) Grant-in-Aid for Scientific Research (B) Grant Number 16H04646, as well as JST CREST Grant Number JPMJCR18J4.

**Author Contributions**

EFT and KI conceived the idea; TO, TM, KI and HI produced the samples and fabricated transport current bridges; NMS and SCW developed the critical current measurement system, which SCW and EFT modified for this study; EFT performed transport current measurements, processed and analyzed the experimental data, and drew the conclusion regarding $p$-wave pairing; WPC developed a full temperature range $p$-wave fitting algorithm and together with EFT performed data fittings; EFT, KI, HI and WPC discussed the results; EFT drafted the manuscript that was edited by WPC, SCW, KI, and HI.

**Competing Interests statement**

The authors declare no competing interests.

# Supplementary Information

**_p_-wave superconductivity in iron-based superconductors**

E. F. Talantsev[1,2,*], K. Iida[3,4], T. Ohmura[3], T. Matsumoto[4], W. P. Crump[5,6], N. M. Strickland[5],

S. C. Wimbush[5,6] and H. Ikuta[3,4]

[1] M. N. Mikheev Institute of Metal Physics, Ural Branch, Russian Academy of Sciences,
18 S. Kovalevskoy St., Ekaterinburg 620108, Russia

[2] NANOTECH Centre, Ural Federal University, 19 Mira St., Ekaterinburg 620002, Russia

[3] Department of Crystalline Materials Science, Nagoya University, Chikusa-ku, Nagoya 464-8603, Japan

[4] Department of Materials Physics, Nagoya University, Chikusa-ku, Nagoya 464-8603, Japan

[5] Robinson Research Institute, Victoria University of Wellington, 69 Gracefield Road, Lower Hutt 5010, New Zealand

[6] MacDiarmid Institute for Advanced Materials and Nanotechnology, PO Box 33436, Lower Hutt 5046, New Zealand

*Corresponding author: E-mail: evgeny.talantsev@imp.uran.ru

**Supplementary Table I.** BCS weak-coupling limit values for $2\Delta(0)/k_BT_c$ and for $\Delta C/C$ and low-temperature asymptotes for the superfluid density, $\rho_s(T)$, for *s*-, *d*-, and *p*-wave pairing [1-5]. For hybrid states the power law exponents were deduced by fittings of the calculated curves of Gross-Alltag *et al.* [2] to the given function, where A and B were free fitting parameters of the order of unity; $k_B$ is the Boltzmann constant; $\Delta_m(0)$ is the maximum amplitude of the *k*-dependent *d*-wave gap, $\Delta(\theta) = \Delta_m(0)\cos(2\theta)$; $\varsigma(3) = 1.2020$ is Riemann's zeta function.

| Pairing symmetry and experiment geometry | $\dfrac{2\Delta(0)}{k_B T_c}$ | $\dfrac{\Delta C}{C}$ | $\rho_s(T)$ low-*T* asymptote |
|---|---|---|---|
| *s*-wave | 3.53 | 1.43 | $1 - 2\sqrt{\dfrac{\pi\Delta(0)}{k_B T}}\, e^{-\dfrac{\Delta(0)}{k_B \cdot T}}$ |
| *d*-wave | 4.28 | 0.995 | $1 - 2\dfrac{k_B \cdot T}{\Delta_m(0)}$ |
| *p*-wave; polar $\mathbf{A}\perp\mathbf{l}$ | 4.92 | 0.792 | $1 - \dfrac{3\pi\ln(2)}{2}\left(\dfrac{k_B T}{\Delta(0)}\right)^1$ |
| *p*-wave; polar $\mathbf{A}\|\mathbf{l}$ | 4.92- | 0.792 | $1 - \dfrac{27\pi\varsigma(3)}{4}\left(\dfrac{k_B T}{\Delta(0)}\right)^3$ |
| *p*-wave; axial $\mathbf{A}\perp\mathbf{l}$ | 4.06 | 1.19 | $1 - \dfrac{7\pi}{15}\left(\dfrac{k_B T}{\Delta(0)}\right)^4$ |
| *p*-wave; axial $\mathbf{A}\|\mathbf{l}$ | 4.06 | 1.19 | $1 - \pi^2\left(\dfrac{k_B T}{\Delta(0)}\right)^2$ |
| *p*-wave; hybrid $\mathbf{A}\perp\mathbf{l}$ | 4.22 | 0.998 | $1 - A\left(\dfrac{k_B T}{\Delta(0)}\right)^{1.2}$ |
| *p*-wave; hybrid $\mathbf{A}\|\mathbf{l}$ | 4.22 | 0.998 | $1 - B\left(\dfrac{k_B T}{\Delta(0)}\right)^{2.6}$ |

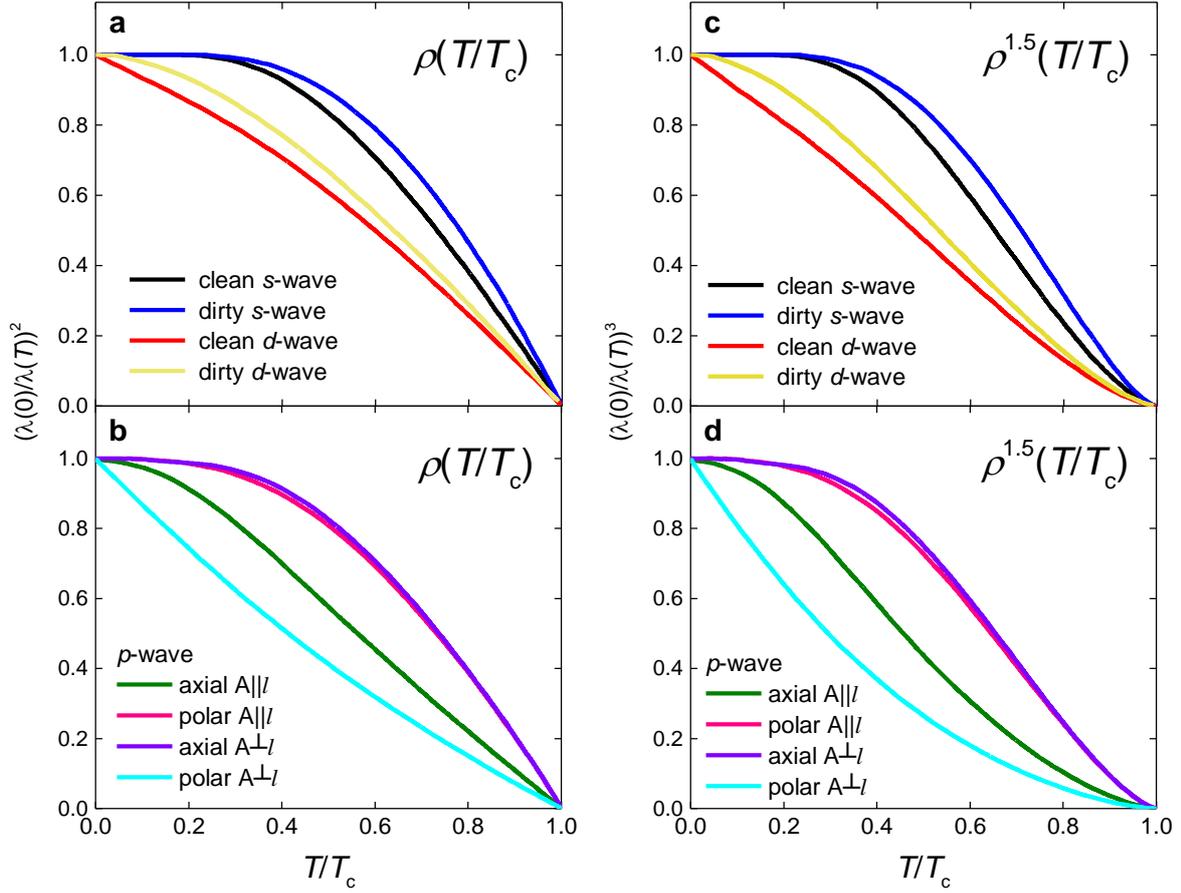

**Supplementary Figure 1.** Calculated normalized superfluid density $\rho_s(T/T_c)$ for (a) $s$- and $d$-wave superconductors, and (b) $p$-wave superconductors. The same results plotted as $\rho_s^{1.5}(T/T_c)$ for (c) $s$- and $d$-wave superconductors, and (d) $p$-wave superconductors.

**Superfluid density measurements of bulk samples**

Below we demonstrate that the temperature-dependent superfluid density, $\rho_s(T)$, measured using muon-spin rotation (μSR) spectroscopy for most iron-based superconductors reported in the literature is also consistent with *p*-wave pairing symmetry in these materials.

***(Li$_{0.84}$Fe$_{0.16}$)OHFe$_{0.98}$Se single crystals.*** In Fig. S2 we show the experimental data for $\rho_{s,ab}(T)$ (the superfluid density in the *a-b* plane) reported by Khasanov *et al.* [6] and a fit using the axial $\mathbf{A}\perp l$ *p*-wave model. There were not enough raw data points near $T_c$, and thus to increase the accuracy of the derived parameters, we reduced the number of free parameters by fixing $T_c$ to the last experimental data point.

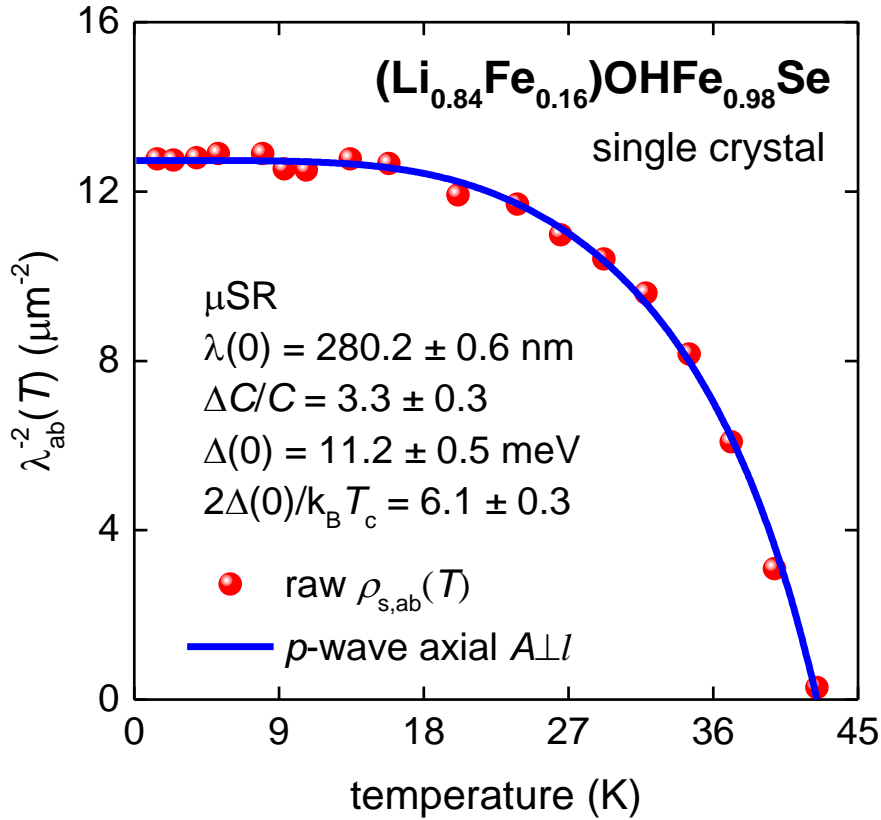

**Supplementary Figure 2.** BCS fits to the experimental $\rho_{s,ab}(T)$ data for a (Li$_{0.84}$Fe$_{0.16}$)OHFe$_{0.98}$Se single crystal [6] assuming a *p*-wave axial $\mathbf{A}\perp l$ model. Derived parameters are: $T_c = 42.5$ K (fixed), $\Delta(0) = 11.2 \pm 0.5$ meV, $\Delta C/C = 3.3 \pm 0.3$, $\lambda(0) = 280.2 \pm 0.6$ nm, $2\Delta(0)/k_B T_c = 6.1 \pm 0.3$. Fit quality is $R = 0.9897$.

As can be seen, $\Delta(0)$ and $2\Delta(0)/k_BT_c$ values deduced from both $J_c(sf,T)$ and μSR data are in excellent agreement with each other, and both indicate moderately strong coupling with *p*-wave gap symmetry in the (Li,Fe)OHFeSe superconductor.

***Rb$_{0.77}$Fe$_{1.61}$Se$_2$ single crystals.*** We examine next the μSR measurements of single crystal Rb$_{0.77}$Fe$_{1.61}$Se$_2$ reported by Shermadini *et al.* [7]. In Fig. S3 we show the raw $\rho_{s,ab}(T)$ data with a fit using the *p*-wave axial $\mathbf{A}\perp l$ model, where again to increase the accuracy of the deduced parameters we fixed the $T_c$ to the last experimental data point.

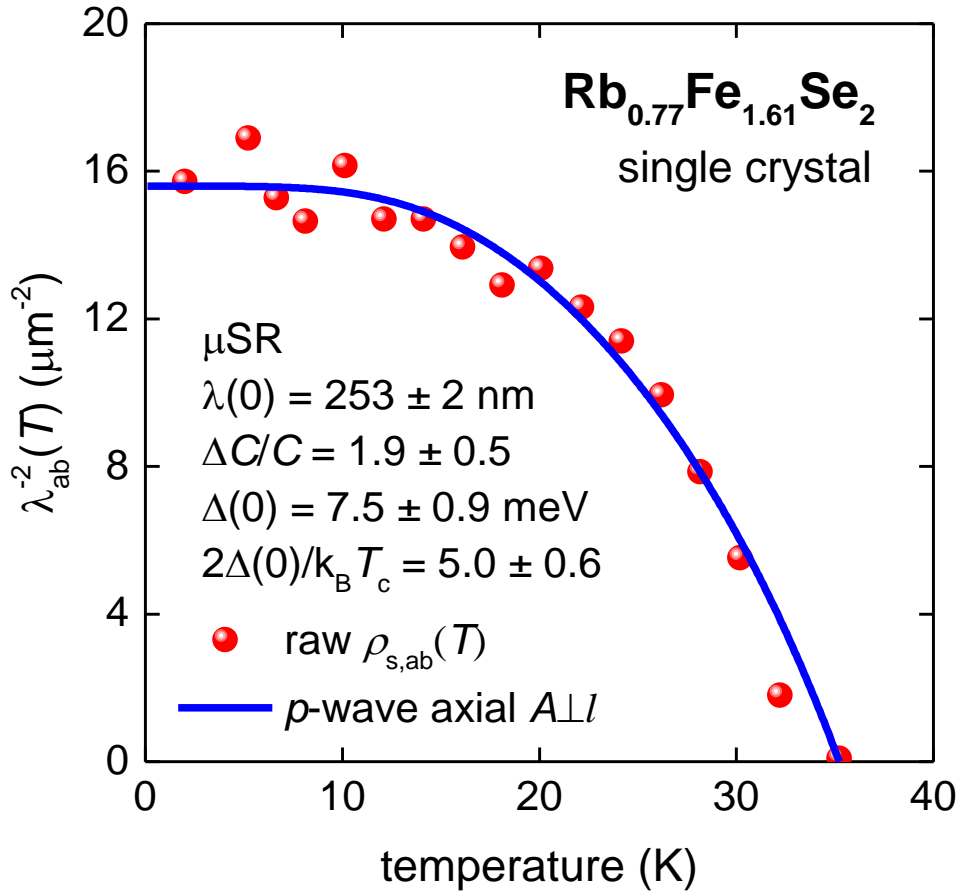

**Supplementary Figure 3.** BCS fits to the experimental $\rho_{s,ab}(T)$ data for a Rb$_{0.77}$Fe$_{1.61}$Se$_2$ sample [7] assuming a *p*-wave axial $\mathbf{A}\perp l$ model. Derived parameters are: $T_c = 35.22$ K (fixed), $\Delta(0) = 7.5 \pm 0.9$ meV, $\Delta C/C = 1.9 \pm 0.5$, $\lambda(0) = 253 \pm 2$ nm, $2\Delta(0)/k_BT_c = 5.0 \pm 0.6$. Fit quality is $R = 0.8764$.

We note that Shermadini *et al.* [7] fitted their data to an *s*-wave model and they deduced a very similar value for $\Delta(0) = 7.7$ meV, which can be compared with our value of $\Delta(0) = 7.5 \pm 0.9$ meV. However, the fit to an *s*-wave model has an unavoidable problem which is the value for the ratio $2\Delta(0)/k_BT_c = 5.5$. This is unrealistically large compared with all other known *s*-wave superconductors [8], and especially the weak-coupling limit of BCS theory of 3.53. By way of comparison, Pb which is a strongly-coupled *s*-wave superconductor has $2\Delta(0)/k_BT_c = 4.86$ [9].

Despite the fact that our ratio is essentially the same $2\Delta(0)/k_BT_c = 5.0 \pm 0.6$, we need to stress that the weak-coupling value for this *p*-wave axial $\mathbf{A}\perp\mathbf{l}$ case is $2\Delta(0)/k_BT_c = 4.06$, which places Rb$_{0.77}$Fe$_{1.61}$Se$_2$ as a moderately strongly coupled superconductor. In addition, the deduced $\Delta C/C = 1.9 \pm 0.5$ is not too far from the weak-coupling limit of $\Delta C/C = 1.2$.

***K*$_{0.74}$Fe$_{1.64}$Se$_2$ *single crystals.*** Shermadini *et al.* [7] also studied in the same paper another iron-based superconductor K$_{0.74}$Fe$_{1.64}$Se$_2$. In Fig. S4 we show the raw $\rho_{s,ab}(T)$ data and a fit using the same *p*-wave axial $\mathbf{A}\perp\mathbf{l}$ model, where again the $T_c$ was fixed to a rounded value close to the last experimental data point.

As was the case for Rb$_{0.77}$Fe$_{1.61}$Se$_2$, Shermadini *et al.* [7] also fitted their data to an *s*-wave model and found $\Delta(0) = 6.3$ meV. In our case we also found this same value ($\Delta(0) = 6.3 \pm 0.4$ meV). However, again for an *s*-wave model the ratio $2\Delta(0)/k_BT_c = 4.7$ is large compared with the majority of other known *s*-wave superconductors [9]. In the *p*-wave case $2\Delta(0)/k_BT_c = 4.5 \pm 0.3$ is in good agreement with a moderately strong coupling pairing strength.

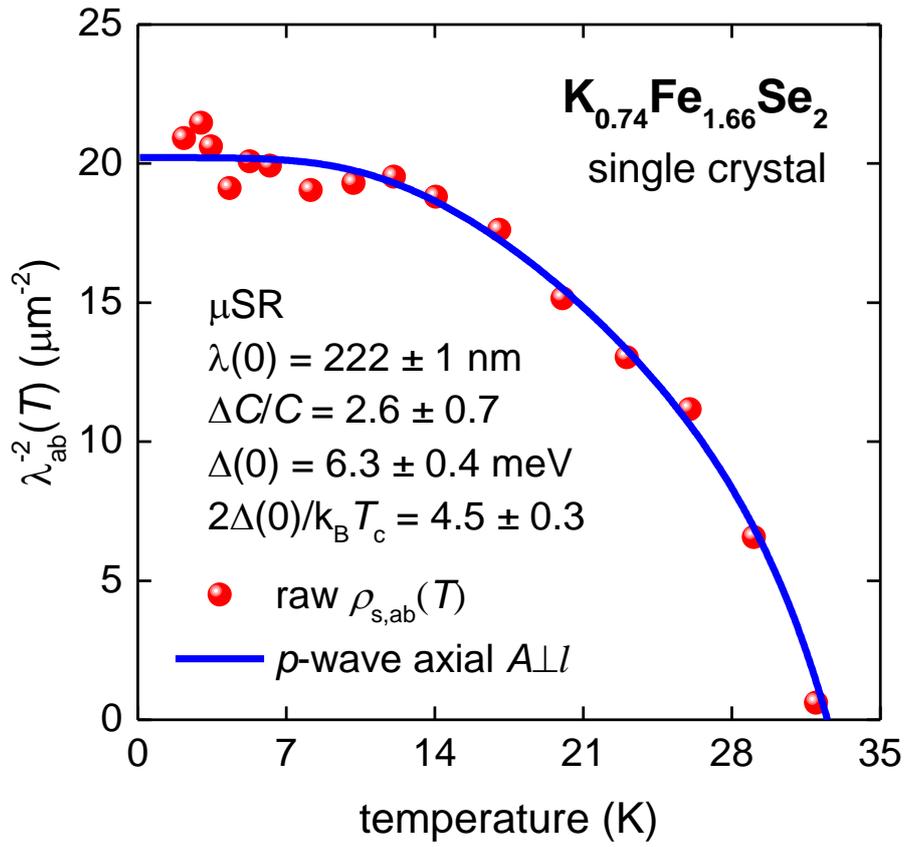

**Supplementary Figure 4.** BCS fits to the experimental $\rho_{s,ab}(T)$ data for a $K_{0.74}Fe_{1.66}Se_2$ sample [7] assuming a $p$-wave axial $\mathbf{A}\perp\mathbf{l}$ model. Derived parameters are: $T_c = 32.5$ K (fixed), $\Delta(0) = 6.3 \pm 0.4$ meV, $\Delta C/C = 2.6 \pm 0.7$, $\lambda(0) = 222 \pm 1$ nm, $2\Delta(0)/k_BT_c = 4.5 \pm 0.3$. Fit quality is $R = 0.8981$.